\newcommand{\be}{\begin{equation}}
\newcommand{\ee}{\end{equation}}
\newcommand{\beeq}{\begin{eqnarray}}
\newcommand{\eeeq}{\end{eqnarray}}

\def\pbar{\bar p}

\def\teta2{\frac {\theta}{2}}
\def\qpri{q^\prime_{\bot i}}
\def\qpritwo{q^{\prime 2}_{\bot i}}
\def\qprii{q^\prime_{\bot {i-1}}}

\def\kti{k_{\bot i}}
\def\kti1{k_{\bot i-1}}
\documentstyle[11pt,epsf,epsfig]{article}

\newlength{\dinwidth}
\newlength{\dinmargin}
\begin{document}

\begin{flushright}
DTP/97/110\\
December 1997
\end{flushright}

\vskip 3cm

\begin{center}
{\large \bf QCD coherence  in deep inelastic scattering
at small $x$ at HERA}\\
\vskip 1.0cm
K. Golec--Biernat\footnote{On leave from  
Institute of Nuclear Physics,
Krak\'ow, Poland; e-mail: K.J.Golec@durham.ac.uk}\\
{\it 
Department of Physics, University of Durham, Durham DH1 3LE, England}
\vskip0.1cm
{L. Goerlich\footnote{e-mail:goerlich@chall.ifj.edu.pl} 
{ and} 
J. Turnau\footnote{e-mail:turnau@chall.ifj.edu.pl}} \\
{\it 
Institute of Nuclear Physics,}
{\it ul. Kawiory 26a, 30-055 Krak\'ow, Poland}

\vskip1cm
\end{center}
\vskip1cm 
\begin{abstract}
QCD coherence 
 effects in initial state radiation at small $x$
in deep inelastic scattering in HERA kinematics
are studied with the help of the Monte Carlo model SMALLX. 
Theoretical assumptions based on the CCFM evolution equation
are reviewed and the basic properties of the partonic final states
are investigated. The results are compared with  those obtained in
the conventional DGLAP evolution scheme.
\end{abstract}

\setcounter{footnote}{0}
\setcounter{page} {0}
\thispagestyle{empty}
\newpage

\section {Introduction}

HERA is the first accelerator which allows one  to study experimentally
the  region  of low values of the Bjorken variable $x$ 
in  deep inelastic lepton-proton 
scattering $(x\sim 10^{-5}$ and $Q^2 > 1~{\rm GeV}^2)$. It is, therefore,
particularly important to analyze experimental 
results in this kinematical region
with the help of  theoretical models which contain  essential
features of QCD in the small-$x$ limit. 

Angular ordering ({\bf coherence}) in initial state gluon radiation (ISR), 
imposed on  real and virtual gluon emission at small $x$,
is one of the most important features of perturbative QCD in the small-$x$ limit
\cite{CIAF,CCFM,CCFM1}. The dominant contribution to ISR is given
by multi-gluon amplitudes with angular ordering of subsequent gluon emissions.
Outside  the angular ordered regions of phase space
destructive interference takes place and amplitudes cancel. This phenomenon
can be formulated in a 
probabilistic framework as a  branching process, 
in which 
the gluon structure function at small $x$ is obtained after a summation
of large $\log(1/x)$ powers  
corresponding to  infrared and collinear 
singularities coming from the angular ordered regions of phase space.
The resummation is effectively 
done with the help of the CCFM evolution equation
\cite{CCFM,CCFM1}.

The  $\log(1/x)$ terms resulting from
angular ordering are subleading with respect to  large logarithms 
obtained in the  
multi-Regge kinematics\footnote{In this kinematics emitted gluons
are strongly ordered in rapidity and have comparable transverse momenta.}
leading to the BFKL  evolution equation \cite{BFKL}. 
In a fully inclusive quantity like 
the gluon structure function the subleading logarithms cancel, and
the CCFM and BFKL evolution equation are equivalent.
This is not true, however,
for exclusive processes for which the ``{\it angular}'' logarithms
give important contribution to the final state structure \cite{CCFM1}. 
Therefore, the branching process with
angular ordering is indispensable for 
the final state description.

SMALLX  is the Monte Carlo  model \cite{SMALLX1,SMALLX2} which incorporates 
the CCFM branching scheme in the initial state gluon radiation. 
For an  alternative Monte Carlo model which also uses the CCFM branching
see \cite{LUNDT}.
The dominant process
at small $x$ -- the quark-antiquark pair production  accompanied by 
gluon radiation, see Fig.1, -- is generated using  
two approximations  for gluon emission.
The  first approximation, 
called {\bf all-loop} and given by the CCFM branching scheme with
angular ordering for all $x$ values, 
is compared   with a conventional branching, called {\bf one-loop}, 
leading to the DGLAP evolution equations with angular ordering at
large $x$ and transverse momenta ordering for small-$x$ values 
\cite{DGLAP,LARGCOH}.
The two branching schemes coincide at large $x$ but
differ significantly when  $x$ is small. 
The cross section for the discussed process is obtained using the 
$k_{\bot}$-factorization theorem \cite{KTFAC}, in  which the off-shell
boson-gluon fusion cross section is convoluted with the gluon
structure function computed in   the  two branching schemes.

The aim of this paper is the  analysis
of angular ordering effects in DIS in the small-$x$ HERA
kinematics with the help of the SMALLX Monte Carlo program. 
The comparison between the CCFM (all-loop) and DGLAP (one-loop) branching
schemes plays an important role in this analysis. The main idea is to
fix the parameters of generation in both approximations seperately such that
a reasonable agreement with the measured inclusive cross section is obtained.
We choose the charm structure function data from HERA for this purpose.
Then, we study various exclusive final state characteristics  to show the
effect of angular ordering. This is done in accord with the 
present belief that only exclusive measurements can discriminate between
mechanisms of QCD evolution.

The paper is  organized as follows. In   section  2 we
briefly describe  
the process under consideration. This includes
discussion of the process kinematics, gluon branching schemes and
their relation to the CCFM and DGLAP evolution equations. In section 3 
we describe
the details of Monte Carlo generation pointing out the crucial elements
discussed later in  section 4 devoted to the
results of our Monte Carlo studies. 

\section {Description of the process}
\subsection{Cross section and kinematics}
The process generated in SMALLX, dominating electron-proton DIS
at small value of the Bjorken variable $x_B$, is shown
in Fig.~1. A quark-antiquark pair is produced accompanied by 
initial state  gluon radiation. 

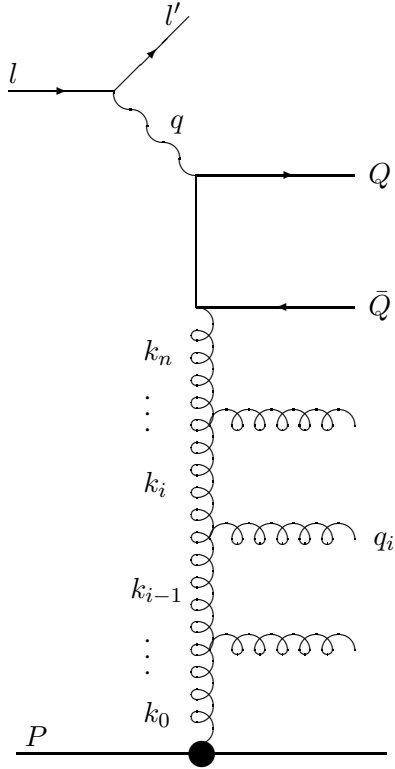
\begin{figure}
\input feynman
\begin{picture}(20000,28500)

\drawline\photon[\NW\REG](14000,22000)[5]
\drawline\fermion[\E\REG](14000,22000)[6000]
\drawline\fermion[\S\REG](\fermionfrontx,\fermionfronty)[5000]
\drawline\fermion[\E\REG](14000,17000)[6000]

\drawarrow[\E\ATTIP](17700,22000)
\drawarrow[\W\ATTIP](17000,17010)
%
\put(20500,21700){$Q$}
\put(20500,16700){$\bar Q$}
\put(13000,23800){$q$}

\drawline\gluon[\S\CENTRAL](14000,17000)[18]
\drawline\gluon[\E\CENTRAL](14500,12500)[5]
\drawline\gluon[\E\CENTRAL](14500,8200)[5]
\drawline\gluon[\E\CENTRAL](14500,4000)[5]
%
\put(12000,10000){$k_i$}
\put(11500,6000){$k_{i-1}$}
\put(20700,8000){$q_i$}
\put(12000,15000){$k_n$}
\put(12000,1300){$k_0$}
%
%
\put(12000,2900){$\cdot$}
\put(12000,3500){$\cdot$}
\put(12000,4100){$\cdot$}
\put(12000,12700){$\cdot$}
\put(12000,12100){$\cdot$}
\put(12000,13300){$\cdot$}

\put(14200,100){\circle*{1000}}

\drawline\fermion[\W\REG](10900,25200)[4000]
\drawline\fermion[\NE\REG](10900,25200)[4000]
\drawarrow[\E\ATTIP](9100,25200)
\drawarrow[\NE\ATTIP](12500,26800)

\THICKLINES
\drawline\fermion[\W\REG](14200,100)[7000]
\drawline\fermion[\E\REG](14200,100)[7000]
\put(7500,400){$P$} 

\put(6900,25500){$l$} 
\put(12800,27700){$l^{\prime}$}

\end{picture}
\caption {The process generated in  SMALLX.}
\end{figure} 

In the high energy limit $\sqrt{s} \gg M$, where $M$ is a quark mass, 
the cross section for this process is given by a convolution
of the hard subprocess $\gamma g \rightarrow Q {\bar Q}$ cross
section ${\hat \sigma}$, and the unintegrated 
gluon structure function ${\cal F}$ describing gluonic emission
\cite{KTFAC}
\be
\label{CS}
\sigma = \frac{\alpha_{em}}{2 \pi} \int \frac{dx}{x} \frac{dy}{y}
\frac{dq^2}{q^2} d^2 {\bf k_{\bot}}
~\bigl \{1+(1-y)^2 \bigr \}
~{\cal F}(x,{\bf k}_{\bot},\mu)
~{\hat \sigma}(M; k, q)~,
\ee
where $y$ is the standard DIS kinematical variable,
$q \simeq y l + q_\bot$ and 
$k=k_n \simeq x_n P + k_\bot$ are the virtual photon and  the last exchanged 
gluon momenta
respectively ($l$ and $P$ are the incident lepton and proton momenta),
and $\mu$ is  a hard scale to be discussed later. 
It is important to stress that ${\hat \sigma}$ is the {\bf off-shell}
cross section, 
which means that the virtuality of the gluon, $k^2=k_\bot^2=
-{\bf k}_\bot^2$, is different from zero. This  allows 
a smooth limit of ${\hat \sigma}$ when a quark mass $M \rightarrow 0$
since the gluon virtulaty $k_\bot^2$ regulates a potential
divergence present in the {\bf on-shell} version of this cross section,
see \cite{SMALLX1,SMALLX2} for more details.

The whole information about the initial state gluon emission
is contained in the unintegrated gluon distribution function
${\cal F}(x,{\bf k}_{\bot},\mu)$ obeying
the CCFM equation \cite{CCFM} which sums  angular ordered gluonic emissions,
for both large and small $x_B$. 
This equation allows  a representation of ${\cal F}$
in terms of branching processes, which is convenient   for
Monte Carlo simulation. 
Before discussing this representation 
it is necessary to describe kinematics of the gluonic emission.

The momenta of exchanged space-like gluons $k_i$, and final state real
gluons $q_i=k_{i-1}-k_i$, are decomposed into a longitudinal and transverse
components (Sudakov decomposition) 
\be
\label{ki}
k_i=x_i p + {\bar x}_i {\bar p}+k_{\bot i}~,
\ee
where the $p$ and $\bar p$ are null vectors 
($2p\cdot \pbar =s$) and $k_{\bot i}\cdot p = k_{\bot i}\cdot \pbar=0$.
In the frame in which $p$ and $\pbar$ are collinear,
$p\sim (1,0,0,1)$ and $\pbar\sim (1,0,0,-1)$, we have
$k_\bot=(0,{\bf k}_{\bot}, 0)$.

Each emission is characterized by the branching  variable
$z_i<1$, being  the fraction 
of the $k^{+}(=k_0+k_3)$ component of the exchanged momentum $k_{i-1}$
lost due to emission of the gluon $q_i$
\be
\label{zdef}
z_i =\frac{k_i^{+}}{k_{i-1}^{+}}~ = \frac{x_i}{x_{i-1}}~.
\ee
Notice the obvious relation
\be 
\label{zprod}
x_n=z_n~z_{n-1}~...~z_1~x_0~, 
\ee
which implies that the small-$x$ limit of the branching process 
is not necessary related to  small values of the   branching variables.
In order to characterize the phase space of multi-gluon emission 
it is  convenient to define the {\bf rescaled} 
transverse momentum of emitted gluons
\be
\label{newpt}
q^\prime_{\bot i} \equiv \frac{q_{\bot i}}{1 - z_i}=
x_{i-1}~\sqrt{s}~\tan{(\frac{\theta_i}{2})}~,
\ee
where from now $q_{\bot i}=|{\bf q}_{\bot i}|$ is 
the length of the transverse component and 
$\theta_i$ is the emission angle of the 
massless final state  gluon, defined with respect to the 
$z$-axis of the collinear frame. 

 
In the approximation leading
to the CCFM equation ({\bf all-loop} approximation) the dominant contribution 
to the multi-gluon emission process comes from a part of 
phase space with {\bf angular ordering} of subsequent emissions
\be
\label{allord}
\theta_i>\theta_{i-1}~~~~~~~~\Leftrightarrow~~~~~~~~
q^\prime_{\bot i} > z_{i-1}~q^\prime_{\bot i-1},
\ee
where the equivalence
results from relations (\ref{newpt}). 
We additionally introduce the {\bf collinear
cutoff} $q_0\equiv z_0 q^\prime_{\bot 0}$ for the first real gluon emission:
$q^\prime_{\bot 1} > q_0$.
Outside of the angular ordered
region destructive interference takes place
and multi-gluon emission amplitudes cancel.
Notice that in the limit $z_i\ll 1$ 
condition (\ref{allord}) practically does not restrict the ordinary
transverse momenta: $q_{\bot i}\approx q^\prime_{\bot i}>0$. 
 
This is not the case in the conventional DGLAP ({\bf one-loop}) approximation,
in which a more restrictive condition is imposed 
\be
\label{oneord}
q^\prime_{\bot i} > q^\prime_{\bot i-1}~,
\ee
and $q^\prime_{\bot 1} > q_0$. Notice that  since 
$q^\prime_{\bot i-1} > z_{i-1}~q^\prime_{\bot i-1}$,
condition (\ref{oneord}) implies angular ordering
for all values of $z$. For $z_i\approx 1$
conditions (\ref{oneord}) and (\ref{allord})
are equivalent, but for $z \rightarrow 0$
the  transverse momentum ordering 
$q_{\bot i} > q_{\bot i-1}$ is additionally 
imposed in the one-loop case on the angular ordered multi-gluon emission.
Therefore, the one-loop phase space
is much smaller than in the all-loop case.

\subsection{Branching probabilities}

Having the momentum $k_{i-1}$ of the exchanged
gluon, the $k_i$ and $q_i$ momenta can be constructed provided 
the branching variable (\ref{zdef}) and the rescaled tranverse momentum
(\ref{newpt}) are known. Therefore, a chain of subsequent gluon
emissions is obtained in which the  branching variables 
$(z_i,q^\prime_{\bot i})$ are generated with the probablity $d {\Pi_i}$, 
given in the all-loop approximation  by
\be
\label{alldistr}
d{\Pi}_i =
\Delta_S(i,i-1)~
{P}_g(i)~
\Theta(\qpri-z_{i-1}\qprii)~
\Theta(1-z_i-{Q_0}/{q^{\prime}_{\bot i}})~ 
\frac{d^2{\bf\qpri}}{{\pi\bf\qpritwo}}~dz_i~
\ee
where $\Delta_S$  and ${P}_g$  are  
functions of branching variables in a shorthand notation
to be discussed below. 
The first $\Theta$ step function imposes angular ordering condition
(\ref{allord}) on generated real gluon momenta, 
whereas the second one allows only emissions with the transverse
momentum $q_{\bot i}=(1-z_i) q^{\prime}_{\bot i} > Q_0$ to  regularize  soft 
singularity in the real gluon emission, manifest at $z_i=1$ in the function
${P}_g$  (eq.(\ref{splfun})).

The function $\Delta_S$, called the Sudakov form factor, sums  virtual gluon
emissions which are also angular ordered. This is reflected in the lower
limit of the first integration  below
\be
\label{sudform}
\Delta_S(\qpri ,z_{i-1} \qprii) = \exp~ \biggl\{  
- \int_{(z_{i-1} \qprii)^2}^{\qpritwo}~
\frac{d{q_{\bot}^{\prime 2}}}{q_{\bot}^{\prime 2}}~
\int_0^{1-Q_0/q_{\bot}^{\prime}} 
\frac{dz}{1-z}~
{\bar \alpha_S} \biggr\}~.
\ee
The strong coupling constant 
$\bar \alpha_S=3\alpha_S/\pi $ depends on 
$q_\bot = (1-z)q_\bot^{\prime}$, and
the upper limit of the $z$ integration 
contains the infrared cutoff $ Q_0$ which now regularize the 
soft  singularity in the virtual emission.
The Sudakov form factor has a simple  interpretation,
being the probability of having no gluon radiation within the
angular region $\theta_{i-1}<\theta<\theta_{i}$.

The function ${P}_g$ is the gluon splitting function 
\be
\label{splfun}
 P_g(z_i, q^2_{\bot i}, k^2_{\bot i}) =
\frac{\bar \alpha_S(q^{2}_{\bot i})}{1-z_i} + 
\frac{\bar \alpha_S(k^{2}_{\bot i})}{z_i}~
\Delta_{NS}(z_i, q^2_{\bot i}, k^2_{\bot i})~,
\ee
where $k_{\bot i}=|q_{\bot 1} + ... +  q_{\bot i}|$
is the total transverse momentum of emitted gluons. The function
$\Delta_{NS}$,  called the non-Sudakov form factor, 
sums virtual corrections relevant only in the
small-$z$ limit,  with infrared singularities due to soft
$(z_i=0)$ exchanged gluons 
\beeq
\label{nonsud}
\nonumber
\Delta_{NS}(z_i, q^2_{\bot i}, k^2_{\bot i}) &=&
\exp~\biggl\{-{\bar \alpha_S(k_{\bot i}^2)} \int_{z_i}^{1} \frac{dz}{z} 
\int {\frac{dk^2}{k^2} \theta(k_{\bot i}-k)~\theta(k-z q_{\bot i})} 
\biggr \}~
\\  \nonumber
\\ 
&=& \exp~\biggl\{-{\bar \alpha_S(k_{\bot i}^2)}~\ln \biggl(\frac{1}{z_i} \biggr)
\ln \biggl(\frac{k_{\bot i}^2}{z_i q^2_{\bot i}} \biggr) \biggr\}~, 
\eeeq
where $k_{\bot i}>q_{\bot i}$ is assumed to get the last equality.
The non-Sudakov form factor screens the $1/z$ singularity
in (\ref{splfun}), suppressing radiation
both for  small  $z$ and $q_{\bot i}\ll k_{\bot i}/\sqrt{z_i}$.

In the {\bf one-loop} approximation the phase space ordering
condition (\ref{oneord}) implies that 
$z_{i-1}=1$ both in the first step function
in (\ref{alldistr}) and  in the Sudakov form factor 
(\ref{sudform}).
In addition, the non-Sudakov form factor $\Delta_{NS}=1$ since
the small-$z$ virtual corrections are relevant 
only in the all-loop approximation.

\subsection{Gluon structure function}

The unintegrated
gluon structure function ${\cal F}(x,{\bf k}_{\bot},\mu)$, which
obeys the CCFM equation \cite{CCFM,CCFM1},  
describes the probablity of finding a gluon with a longitudinal
momentum fraction $x$ and transverse momentum ${\bf k}_{\bot}$
at the hard scale $\mu$, being related to the maximal angle of the 
emitted gluon.
The additional dependence
of ${\cal F}$ on the hard scale is a particular feature of the CCFM equation,
in contrast to  the BFKL equation that, being conformally invariant, does not
introduce an additional scale into the unintegrated gluon structure function.

The CCFM equation allows one to represent ${\cal F}$ as a branching process
with the help of the branching probabilities (\ref{alldistr})
\beeq
\label{strfun}
&~&{\cal F}(x,{\bf k}_{\bot},\mu,q_0) = 
\delta(x-x_0)~\delta^2({\bf k}_{\bot}-{\bf k}_{\bot 0})~
\Delta_S(\mu,q_0)~
\Theta(\mu-q_0) + 
\\ \nonumber
\\ \nonumber
&~&\sum_{n=1}^{\infty}~\int...\int
~\Theta(\mu-z_n q^{\prime}_{\bot n})
~\Delta_S(\mu,z_n q^{\prime}_{\bot n})~
\biggl \{ \prod_{i=1}^{n}
d{\Pi}_i \biggr \}
~\delta(x-x_n)
\delta^2({\bf k}_{\bot}-{\bf k}_{\bot n})~,
\eeeq
where $(x_n,{\bf k}_{\bot n})$ are parameters of 
the last exchanged gluon momentum in the chain: 
$k_n=x_n p+{\bf k}_{\bot n}$, see Fig.~1, and we explicitly indicate
the presence of the collinear cutoff $q_0$ for the first real emission.

The integration in eq.~(\ref{strfun}) is performed over
angular ordered phase space of  the real gluon emission 
-- the integration variables
and their range are implicit in the branching probabilities 
$d{\Pi}_i$ (\ref{alldistr}).
The additional $\Theta$ functions 
introduce  a hard scale $\mu$ giving
the upper limit for the phase space integration 
to be discussed in detail in section 3. The soft 
singularities regularized by the cuttof $Q_0$ in $d{\Pi}_i$
cancel  after the phase space integration and the limit 
$Q_0\rightarrow 0$ can be taken for $\cal F$. 
The collinear divergence related
to the first gluon emission and regularized by $q_0$
stays as an important parameter in the branching procedure,
providing the scale for the  non-perturbative  distribution 
${\cal F}^0(x_0,{\bf k}_{\bot 0},q_0)$, 
to be convoluted with ${\cal F}$.

The structure function ${\cal F}$ integrated over 
transverse momentum up to the scale $\mu=Q$ becomes the usual gluon structure
function (gluon density)
\be
\label{sfint}
F(x,Q,q_0) \equiv \int d^2 {\bf k_{\bot}}~{\cal F}(x,{\bf k_{\bot}},Q,q_0)~
\Theta(Q-|{\bf k_{\bot}}|)~.
\ee
We are interested in the $Q^2$ dependence (evolution) of $F$ at small $x$
given by the CCFM equation, where for simplicity 
the  fixed strong coupling constant $\bar \alpha_S$ is assumed.
The result,  given in terms of the moments $F_\omega$ of 
the gluon structure function $F$, is the following
\be
\label{GLUON}
F_\omega(Q,q_0) \equiv \int_0^1 dx~x^{\omega}~F(x,Q,q_0)
= \biggl( \frac {Q^2}{q_0^2} \biggr)^{\gamma({\bar \alpha_s}/{\omega})}~,
\ee
where the gluon anomalous dimension $\gamma$ is equal to that
obtained in the analysis of the BFKL equation
\be
\label{CCFMAD}
\gamma(\frac{\bar \alpha_s}{\omega})=
\gamma^{BFKL}(\frac{\bar \alpha_s}{\omega})={\frac{\bar \alpha_s}{\omega}} 
+ 2 {\zeta_3} {\biggl( {\frac {\bar \alpha_s}{\omega}} \biggr)}^4 
+ 2 {\zeta_5} {\biggl( {\frac {\bar \alpha_s}{\omega}} \biggr)}^6 + ~...~,
\ee
and $\zeta_i$ is the Riemann zeta function. 

It should be noted that the BFKL  anomalous dimension 
was derived 
in the all-loop (CCFM) approximation as a result of
the complete cancellation of large $\log(1/x)$ powers 
coming from the integration
over the angular ordered phase space of the real and virtual emissions. 
These logarithms give 
subleading corrections to the BFKL equation and are fully cancelled in 
the gluon structure function at small $x$.
They are not cancelled, however, for more  exclusive quantities,  
and in such cases
the all-loop approximation
with  angular ordering is essential for the  proper 
description of exclusive processes, 
see  \cite{CCFM1} for recent discussion.
The subleading corrections in the CCFM equation also 
decrease  the BFKL value of the 
QCD Pomeron intercept $\omega_0=(4\ln2) {\bar \alpha_s}$
in the asymptotic formula: ${\cal F}(x) \sim x^{-(1+\omega_0)}$, and  reduce 
diffusion  in the transverse momentum space, see \cite{KW,SAL}.

In summary, the CCFM equation is a generalization of the BFKL equation
by taking into account a part of subleading $\log(1/x)$ corrections
generated by angular ordering. These corrections
cancel in the fully inclusive quantity like 
the gluon structure function 
(\ref{sfint})  but give an important contribution to more 
exclusive quantities (e.g. gluon multiplicity).

In the one-loop  approximation the small-$x$ behaviour is given
by the gluon anomalous dimension equal to the first term in  expansion
(\ref{CCFMAD}): $\gamma({\bar \alpha_s}/{\omega})={\bar \alpha_s}/{\omega}$.
This result is in a perfect agreement with the gluon anomalous dimension
obtained from the DGLAP equation in the small-$x$ limit.

\section{Monte Carlo simulation}

The cross section ({\ref{CS}) and  branching formulae
({\ref{strfun})  form the   basis for the Monte
Carlo generation  in  SMALLX.

The procedure starts with generation of an initial state parton shower,
see Fig.1. In the first step the gluon
momentum $k_0=x_0 P+k_{\bot 0}$ is generated according to a given
input ${\cal F}(x_0,{\bf k}_{\bot 0})$. Then  next gluon emissions
are generated with distributions (\ref{alldistr}). 
At the $i^{th}$ vertex, $q^{\prime}_{\bot i}$  is 
selected with the help of the Sudakov form factor (\ref{sudform}), and 
an azimuthal angle is chosen randomly
to form a vector  ${\bf q}^{\prime}_{\bot i}$. 
The branching fraction $z_i$ is chosen next according to   
distribution   (\ref{splfun}).
Notice that $P_g$ depends on the  variable 
$k_{\bot i} =  |{\bf k}_{\bot i-1}+{\bf q}_{\bot i}|$ thus,  
in order to simplify the generation, the approximation 
${\bf q}_{\bot i} \simeq {\bf q}_{\bot i}^\prime$ is made when  
the selection of $z_i$ is done. Then 
the emitted gluon momentum ${\bf q}_{\bot i}=(1-z_i){\bf q}^{\prime}_{\bot i}$
is computed and the true value of ${\bf k}_{\bot i}$ calculated. 

The branching continues until the stopping condition is reached, i.e.
for certain $n$ we have
\be
\label{allstop}
z_n q^\prime_{\bot n} = x_{n}~\sqrt{s}~\tan ({\theta_{n}}/{2})< \mu~,
\ee
and $q^\prime_{\bot n+1}$ momentum  in the next emission  exeeds the  bound.
In this case the $n$-gluon final state is generated.
In the {\bf all-loop} approximation
the bound $\mu$ results from condition (\ref{allord}) 
\be
\label{angbound}
\theta_1 < ... < \theta_n < \Theta~,
\ee
where $\Theta$ is the emission angle of  
the quark-antiquark  total  momentum 
$p_Q+p_{\bar Q} = Y p +  {\bar Y}  {\bar p} + Q_\bot$. Thus, it is easy to see
that the choice 
\be
\label{bound}
\mu = x_{n}~\sqrt{s}~\tan({\Theta}/{2})~
\ee
gives the angular ordering condition (\ref{angbound}).
In the {\bf one-loop} approximation the stopping condition takes
the form 
\be
\label{onestop}
q^\prime_{\bot n} < \mu=\sqrt{Q^2+M^2}~,
\ee
where $\mu$ is given by the scale at which
the hard cross section $\hat \sigma$ starts to be  strongly suppressed 
as a function of the gluon virtuality $k^2_{\bot}$. We will show 
in the forthcoming analysis that  condition (\ref{onestop}) is much
more restrictive than (\ref{allstop})
 and leads to smaller values of transverse
momenta than in the all-loop case.

The quark-antiquark pair can be produced provided 
\be
\label{spair}
\hat s \equiv (k+q)^2 > 4 M^2~,
\ee
where $k=k_n$ is momentum of the last gluon in the chain and
$q \simeq y l + q_\bot$ is the photon momentum, see Fig.~1, generated
with the distribution $dy/y~dq_{\bot}^2/q_{\bot}^2$.
If condition (\ref{spair}) is satisfied 
the hard cross section $\hat \sigma$ in eq.~(\ref{CS}) is calculated and
the event is accepted with a non-zero weight, otherwise
the event weight is equal to zero.

In general, the non-zero weight is 
a product of subweights coming from the virtual photon state generation,
the gluon branching and  the hard scattering. At each stage of  generation
the subweights could become zero when momenta fall outside
kinematical limits. Condition (\ref{spair}) is quite restrictive 
in this respect and only
a fraction of gluonic events can produce a quark-antiquark pair.

At the end of the generation the longitudinal momentum carried
by gluons is computed at the  scale $\mu$ 
\be
\label{glumom}
P(\mu)= \int_0^1 dx~x F(x,\mu)~,
\ee
where $F$ is given by eq.~(\ref{sfint}). The conservation of $P(\mu)$
is slightly violated when  $\mu$ runs, thus eventually
all weights are renormalized by the factor $0.5/P(\mu)$ to get
gluons carrying half of the proton's momentum.

\section{Monte Carlo results}

We shall present the results of our first experience with the 
SMALLX Monte Carlo program in HERA kinematics
($27.6~{\rm GeV} + 820~{\rm GeV}$ for electron and proton energy, respectively).
Most of the results will be presented in the HERA laboratory frame
in which the $z$-axis is defined by the incoming proton 
direction\footnote{
This frame is also used in SMALLX to define decomposition
(\ref{ki}): $p=P$ and $\bar p=l$, 
where $P$ and $l$ are the colliding proton and lepton momenta, respectively.}.
Some results
are presented in the virtual photon-proton frame (CMS
frame) with the $z$-axis also defined by the  proton direction.

We confine ourselves to the study of  open charm production in DIS at small
$x$, with 
the charm quark mass $M=1.5~{\rm GeV}$. This is motivated by our strategy to fix
the generation parameters, in both approximations
separately,  in order to describe the measured at HERA inclusive quantities 
like structure functions. Then we  compare
more exclusive final state characteristics to understand the difference
between the two gluon branching schemes. Within such an
approach  open charm production is the best process to consider 
since it is not spoiled by the details of the light flavours generation.

We  fix the generation parameters to describe
the charm contribution $F^{c\bar c}_2$ to the proton structure function
$F_2$ measured in DIS processes at HERA \cite{F2CC}.
This includes the parameters of the initial gluon distribution ${\cal F}^0$,
for which the following form is postulated
\be
\label{initglu}
x_0 {\cal F}^0(x_0,{\bf k}_{\bot 0},Q_s)= A_0~x_0^{-p_0}~(1-x_0)^{p_1}~
\exp(-2 {{\bf k}^2_{\bot 0}}/Q_s^2)~,
\ee 
where the parameter $Q_s$ also plays  the role of the collinear cutoff
$q_0$ for the first real gluon emission, and the normalization
constant $A_0$ is determined from the momentum sum rule 
$P(Q_s)=0.5$ (\ref{glumom}). We have fixed  $Q_s=1~{\rm GeV}$ and 
$p_1=5$ motivated by large-$x$ behaviour of the conventional gluon
distribution. For the purpose of Monte Carlo generation we 
keep the non-zero value of the infrared parameter 
$Q_0$ in (\ref{alldistr})
giving the minimal
transverse momenta of emitted gluons: $Q_0=1~{\rm GeV}$.
The  parameter  $p_0$ is chosen to obtain a good
description of $F_2^{c \bar c}$ at $Q^2=12~{\rm GeV}^2$, whereas
the result for higher  values  is given
by the evolution embodied in both approximations.
In order to achieve that, 
the shapes of the initial distributions have to differ significantly:
for the one-loop mode  the small-$x$ values are strongly enhanced 
$(p_0=0.4)$ in comparison to the all-loop  initial distribution
$(p_0=0.0)$, see the corresponding lower curves in Fig~\ref{fig:22}.

In Fig.~\ref{fig:2} we show $F_2^{c \bar c}$ at 
$Q^2=12,~25$ and $45~{\rm GeV}^2$ for all-loop (solid lines)
and one-loop (dotted lines) approximations 
with the parameters defined above.
The evolution does not differ significantly  once
the parameters have been tuned at $Q^2=12~{\rm GeV}^2$, and
both modes  describe $F_2^{c \bar c}$ in the whole measured range
quite well. This result indicates that 
there is no significant difference between the
$Q^2$ evolution in the two approximations when the asymptotic 
$x$-behaviour of $F_2^{c \bar c}$ is obtained.
Notice that in the one-loop approximation the asymptotic behaviour is
imposed by hand, through the steep initial gluon distribution, 
whereas in the all-loop mode is a result of the CCFM 
evolution -- compare
the two corresponding upper curves in Fig.~\ref{fig:22}.
The dashed lines in Fig.~\ref{fig:2}, 
obtained from the flat $(p_0=0.0)$ input distribution, 
illustrate the importance
of the steep form of the initial distribution in the one-loop mode
for the data description.

\subsection{Gluon branching properties}

Having fixed the generation parameters we compare  different characteristics
of the gluon branching in the two approximations for two cases. 
In the first case, we analyze the gluonic state generated until the stopping
condition (\ref{allstop}) or (\ref{onestop}) is satisfied, 
with the event weights
given only by the gluon branching procedure. In the second case, we
take into account only those gluonic events which  satifisfy
condition (\ref{spair}) for  quark pair production, and the full
event weight, with the $\gamma g \rightarrow Q {\bar Q}$ subprocess 
contribution $\hat \sigma$, 
is computed. We analysed the small-$x$ DIS regime,~ 
$1.6\cdot 10^{-4}<x_B<1.3\cdot 10^{-2}$ for 
$<Q^2>=12~{\rm GeV}^2$ and $0.01<y<0.8$,
based on the statistics of $10^7$ generated events.

In Fig.~\ref{fig:5} we show the distribution of the transverse
and longitudinal components of the last exchanged gluon 
momentum $k_n=x_n P+k_{\bot}$, see Fig.~1.
The most striking feature 
of transverse momentum distribution (two upper plots) is a dramatic
difference in the hard part of the spectrum  
($k_{\bot}>5~{\rm GeV}$), where
the hard tail in the all-loop approximation is strongly enhanced 
(solid lines); see also analysis (\cite{SMALLX2}).
This phenomenon can be traced back to  the all-loop
condition (\ref{angbound})  which
imposes only an angular bound on the maximal angle for 
emitted gluons, in contrast to the
one-loop condition (\ref{onestop}) that also bounds their transverse momenta.
Since $k_{\bot}$ is the total transverse momentum of emitted gluons,
$k_{\bot}=|q_{\bot 1}+...+q_{\bot n}|$, this effect is also clearly visible
in the presented spectra. 
There is no difference in the longitudinal momentum spectra
between the two approximation (two lower plots) as a result of the choice
of the input distributions.
 
The events with 
the charm pair produced (``{\it with box}'' plots) have suppressed
small values of  $x_n$  and large values of
$k_{\bot}$  in comparison to the ``{\it no box}'' plots, 
where all generated gluonic events are analysed.
This can be easily understood
if condition (\ref{spair}) is written for $q\approx y l+q_{\bot}$ and
$k\approx x_n P+k_{\bot}$
\be
\frac{x_n}{x_B}> \frac{4M^2+({\bf k}_{\bot}+{\bf q}_{\bot})^2}{Q^2}~.
\ee
Clearly, large values of $x_n$ and small values of $|{\bf k_{\bot}}|$ are
favoured by  quark-antiquark pair production.  Therefore, the  gluon
distribution is dynamically suppressed in the generation 
of the full final state for $x=x_n<10^{-3}$, as is seen in 
the lower ``{\it with box}'' plot in Fig.~\ref{fig:5}; see also
\cite{SMALLX2}}.  

The analysis of the final state real gluons, done in the same spirit,
is presented in Fig.~\ref{fig:4}. 
The two upper plots with the gluon multiplicity
distribution exhibit a pronounced difference between the 
branching schemes. Due to  angular ordering (\ref{allord}) 
there are significantly more gluons in the all-loop scheme, 
and their transverse momenta can be much bigger than in the one-loop
case -- see the plots in the second row, where the hard tail
in the all-loop $q_{\bot}$-spectrum  persists.
Notice that most gluons are ``soft'' ($q_{\bot}\sim 1~{\rm GeV}$) and there
are significantly more ``soft'' gluons in the all-loop approximation.
The sharp edge at $1~{\rm GeV}$ is the result of the infrared cutoff $Q_0$ 
imposed on transverse momenta. 
The  plots in the bottom row show  transverse momentum as a
function of rapidity ($E_{\bot}$ flow). The difference between
the two approximations in these plots
reflects the fact that there are more
gluons (both ``soft'' and ``hard'') in the CCFM branching.

It is interesting to analyse
the role of the hard scale (\ref{bound}) in the all-loop approximation,
related to  the  maximal angle for the gluon emission.
In Fig.~\ref{fig:4stop} we
show the same   characteristics as in Fig.~\ref{fig:4}, 
but now the one-loop
stopping condition (\ref{onestop}) is used in the all-loop generation.
The $q_{\bot}$-spectrum reveals that the scale (\ref{bound}) is
responsible for the hard tail in the all-loop approximation -- the hard tail
disappears when the one-loop bound
(\ref{onestop}) is imposed in the all-loop mode, while the ``soft'' 
part of the spectrum stays intact.
The other all-loop characteristics are not so dramatically affected. 
The average number of gluons 
is smaller and there are less radiation events with more than $n=4$ gluons
in the ``{\it with box}'' case (upper plots).
The significant difference between the generation modes persists in the
$E_{\bot}$ flow (bottom plots) since 
the ``soft'' part of the $q_{\bot}$-spectrum is not affected by the
one-loop stopping condition.  The lack of the ``hard'' gluons
($q_{\bot} > 5 {\rm GeV}$) 
influences mostly the $E_{\bot}$ flow in the central rapidity region
$-1<\eta<1$. In summary, angular ordering (\ref{angbound})
in the all-loop approximation leads to 
significantly more gluons radiated into the final state with the 
$k_{\bot}$-spectrum much harder than in the one-loop approximation.

\subsection{Final state analysis}

In this section we study the full final state properties, including
the charm-anticharm quark pair. Therefore, for each analysed event 
condition (\ref{spair}) is satisfied  and the final weight is computed.
Before presenting the results let us stress that we study
the very nature of the CCFM and DGLAP branching schemes, 
since the additional effects like the final
state radiation from quarks and gluons or hadronization effects
are neglected. These effects remain to be included in SMALLX
to make the comparison with the real exclusive data.

In Fig.~\ref{fig:6} we show two  final state characteristics discussed
in the context of the BFKL signatures in the HERA data: 
the transverse momentum 
spectra \cite{PTSPEC,PTSPECH1} and the tranvserse energy flow \cite{ETFLOW},
for the small-$x$ kinematics specified in the previous section.
The presented plots
reveal how important the charm and anticharm quarks are for the final state
characteristics. The difference between the approximations in the hard
part of the gluonic $q_{\bot}$-spectrum is significantly reduced by the quark
$q_{\bot}$-distribution. 
However, the difference between the approximations persists, 
favouring harder spectrum in  the all-loop   mode which stays
in a qualitative agreement with the recent HERA data \cite{PTSPECH1}.
The $E_{\bot}$ flow is less affected by the quark inclusion, 
especially in the forward region,  
but we expect significant hadronization
corrections in this region. 

In Fig.~\ref{fig:7} 
we show the mean transverse energy flow $\langle E_T \rangle$
in the central region of pseudorapidity in the $\gamma p$ CMS
frame for several values of $x$.  $\langle E_T \rangle$ decreases
 with increasing $x$ for the all-loop generation mode,
and increases in the one-loop case. This result is in 
a qualitative agreement with the semi-analytical calculations done in
\cite{ETFLOW}, which are based on the BFKL and DGLAP equations.

The diffusion of the gluon transverse momenta $k_{\bot}$ 
along the gluon chain as $x_n$ decreases
can result in large values of $k_{\bot}$ in the all-loop approximation,
see Fig.~\ref{fig:5}.
This phenomenon leads to azimuthal decorrelation
between the quark and antiquark momenta \cite{ANGCOR}. 
In Fig.~\ref{fig:8} we show
the distribution of the difference $\Delta\phi$ between the
azimuthal angles of the quark and antiquark momenta,
calculated in the $\gamma p$ CMS frame, for two different values of 
the Bjorken variable $x_B$. 
In the all-loop mode the distribution shows weakening of the azimuthal
(back-to-back) correlation as $x$ decreases, 
while for the one-loop case this effect is not observed, and
the distribution is concentrated around $\Delta\phi=\pi$,
independent  of $x$ values. 

In Fig.~\ref{fig:9}  
we present another widely advertised signature of the
BFKL dynamics at HERA: the cross section for
forward jet production \cite{FOJET}, where
the jets were chosen as
in the H1 analysis \cite{FOJETH1}. In this process a forward jet
with respect to the proton direction with transverse momentum
$q_{\bot}\sim Q$ is measured, where $Q$ is the virtuality of the photon.
The cross section for such a measurement
is significantly enhanced as
a result of gluonic emissions
populating the rapidity region between the forward jet and the 
virtual photon. 
The original proposition \cite{FOJET} concerns the gluonic emission in
the Regge kinematics, but in view of discussion in \cite{CCFM1} the angular
ordered kinematics imposed in the CCFM scheme is more appropriate.
We remind the reader that
all our results refer to  charm production, so
that the presented forward jet 
cross sections are much smaller than those presented in \cite{FOJETH1}.

The total forward jet cross sections, shown  in the upper 
plot of Fig.~\ref{fig:9}
as a function 
of $x_B$ for the two generation modes, 
differ significantly.  
The all-loop cross section is bigger and much steeper than in
the one-loop approximation. The two lower plots
with  the  pure gluonic and pure quark forward jet cross sections
explain the reason  for the difference seen in the upper plot.
In the all-loop approximation the total cross section is
dominated by the gluonic forward jet contribution and the pure quark component
is almost negligible for the smallest $x_B$ values, whereas
in the one-loop mode the  gluon and quark contributions are comparable, and
much smaller than the gluonic cross section in the all-loop case.
Clearly, forward jet production is a good process to discriminate
between the two gluon branching schemes.

\section {Conclusions and outlook}
We have presented Monte Carlo studies of angular ordering effects
in initial state gluon radiation in the small-$x$ HERA kinematics, 
described by the CCFM (all-loop) branching scheme.
The studies have shown the relevance of the CCFM approximation
in the description of exclusive processes in DIS at HERA.
In contrast to the conventional DGLAP (one-loop) approximation the 
CCFM scheme allows more gluons in the final state  with a much broader
transverse momentum spectrum. As we have shown this is  the result of 
the angular ordering condition (\ref{angbound}) imposed on  the phase
space of gluon emissions in the all-loop approximation. 

The gluonic spectra are altered by the quark
pair production but the  angular ordering effects 
are still visible in the presented characteristics.
The most spectacular results are 
the azimuthal angle decorrelation for
the heavy quarks -- a direct consequence of $k_{\bot}$-diffusion along the
gluon chain in the CCFM scheme, and  forward jet production. 
The rest of the results may strongly depend
on such effects like the final state branching, hadronization
or inclusion of quarks in initial state radiation.
Therefore the additional effort is necessary to include these
effects into SMALLX, 
before the  meaningful comparison with HERA data could be done.
We hope to report on progress in this direction in the near future.

\newpage
\centerline{\bf Acknowledgements}

We are grateful to Bryan Webber for providing us with his Monte
Carlo program SMALLX  and for valuable discussions.
Discussions with  Gunnar Ingelman, 
Hannes Jung, Michael Kuhlen, Jan Kwieci\'nski, Sabina Lang, Claire
Lewis, Ewelina \L obodzi\'nska, Giu\-seppe Marchesini, 
Alan Martin and Peter Sutton are gratefully acknowledged.
We also thank  John Campbell  for a careful reading of the manuscript.
K.G thanks the Royal Society for financial support and the Department of
Physics of the University of Durham for its warm hospitality. 
This research has been supported in part by the
Polish State Committee for Scientific Research, grants
no.2 P03B 231 08, 2 P03B 089 13 and 2 P03B 055 13.

\newpage

\begin{figure}[htb]
   \vspace*{-1cm}
    \centerline{
     \psfig{figure=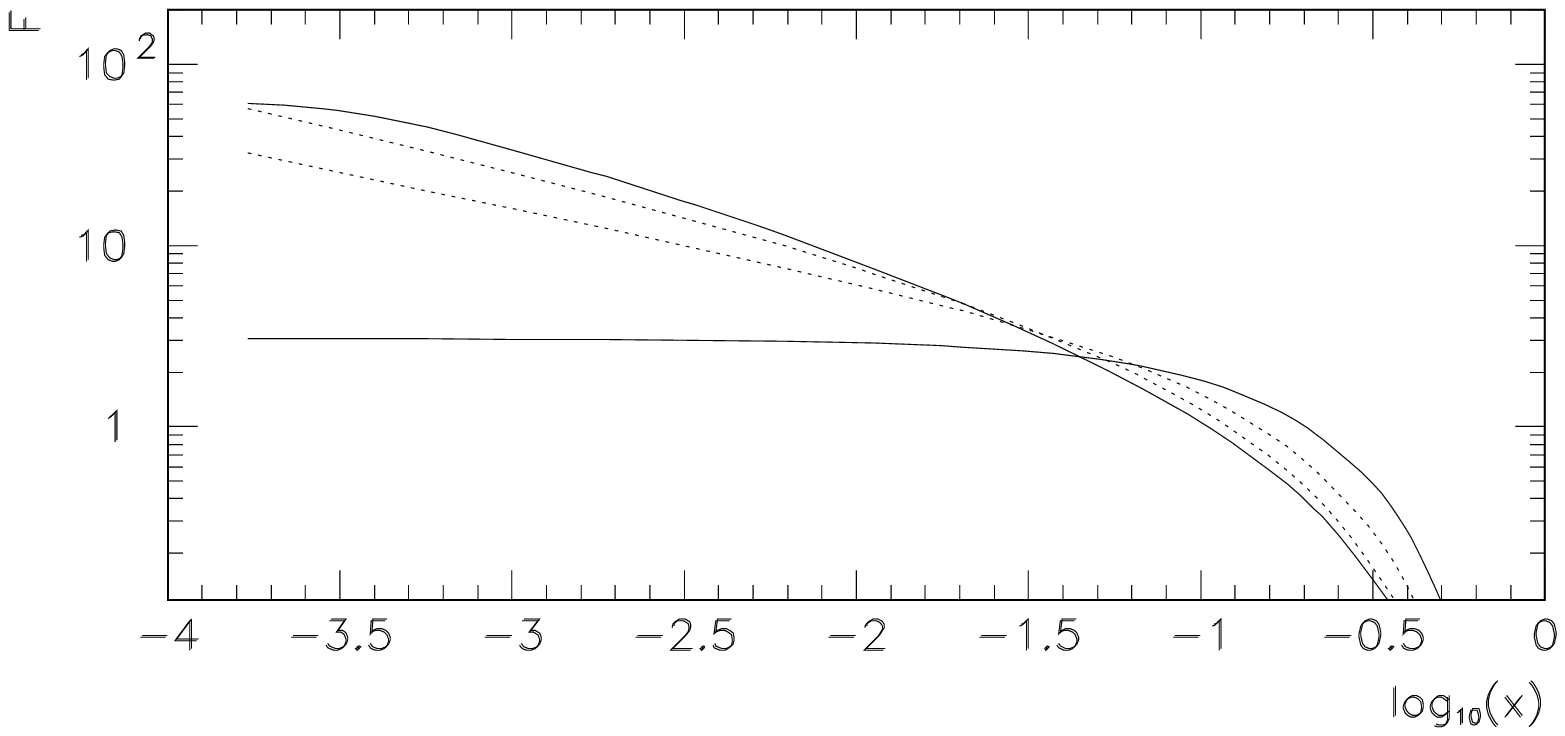,height=10cm,width=18cm}
               }
    \vspace*{-0.5cm}
     \caption{The integrated gluon distribution $F$ in all-loop (solid) and
     one-loop (dotted) schemes at the initial scale $1~{\rm GeV}^2$ 
     (the corresponding
     lower curves),
     and after evolution to $<Q^2>=12~{\rm GeV}^2$ (the
     corresponding upper curves).     
}\label{fig:22}
   \vspace*{0cm}
    \centerline{
     \psfig{figure=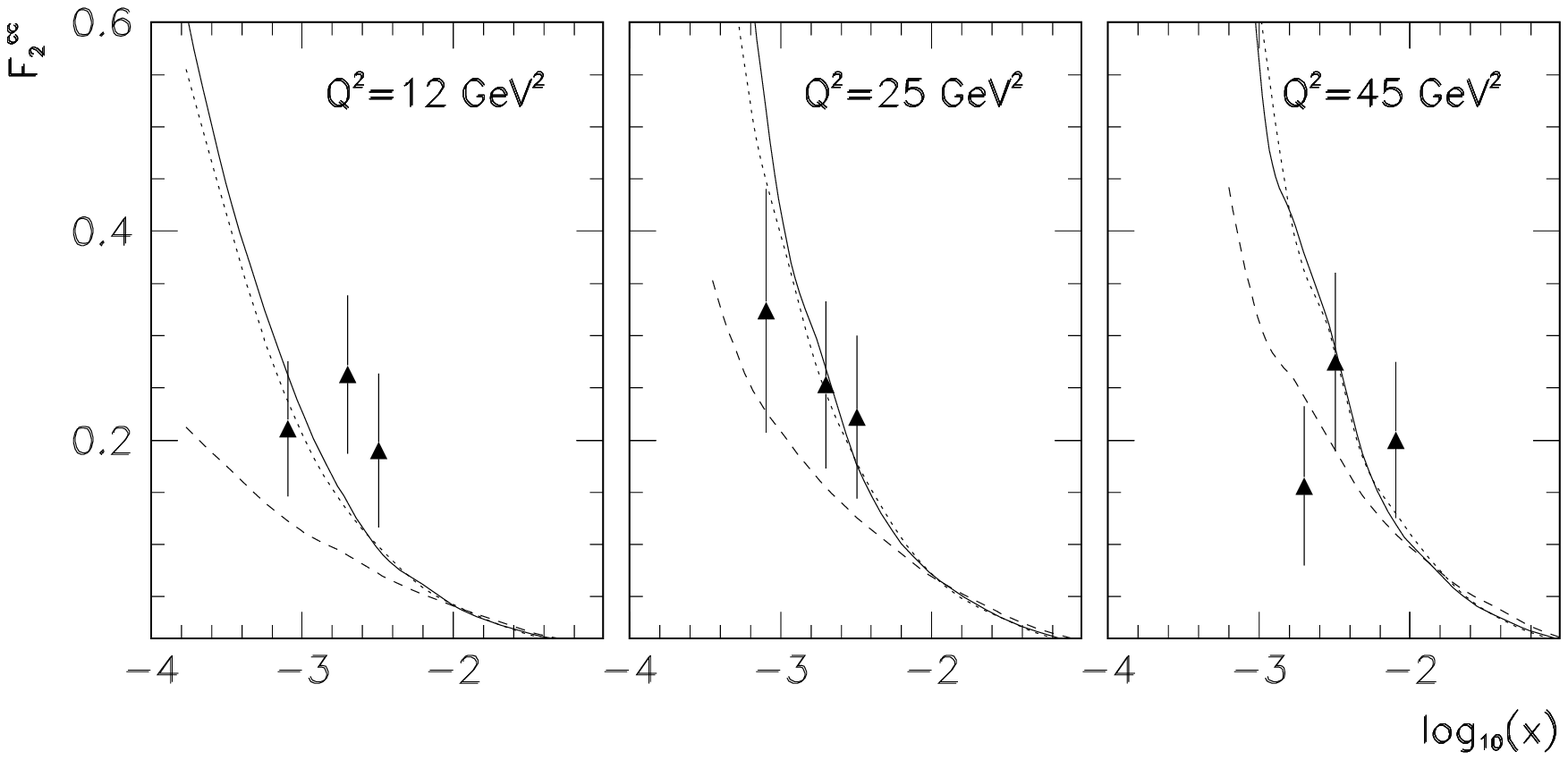,height=10cm,width=18cm}
               }
    \vspace*{-0.5cm}
     \caption{$F_2^{c \bar c}$ in the all-loop (solid lines) and 
one-loop (dotted lines) approximations. The dashed
lines are obtained for the flat input distribution in the one-loop mode.
}\label{fig:2}
\end{figure}
\newpage

\begin{figure}[htb]
   \vspace*{-1cm}
    \centerline{
     \psfig{figure=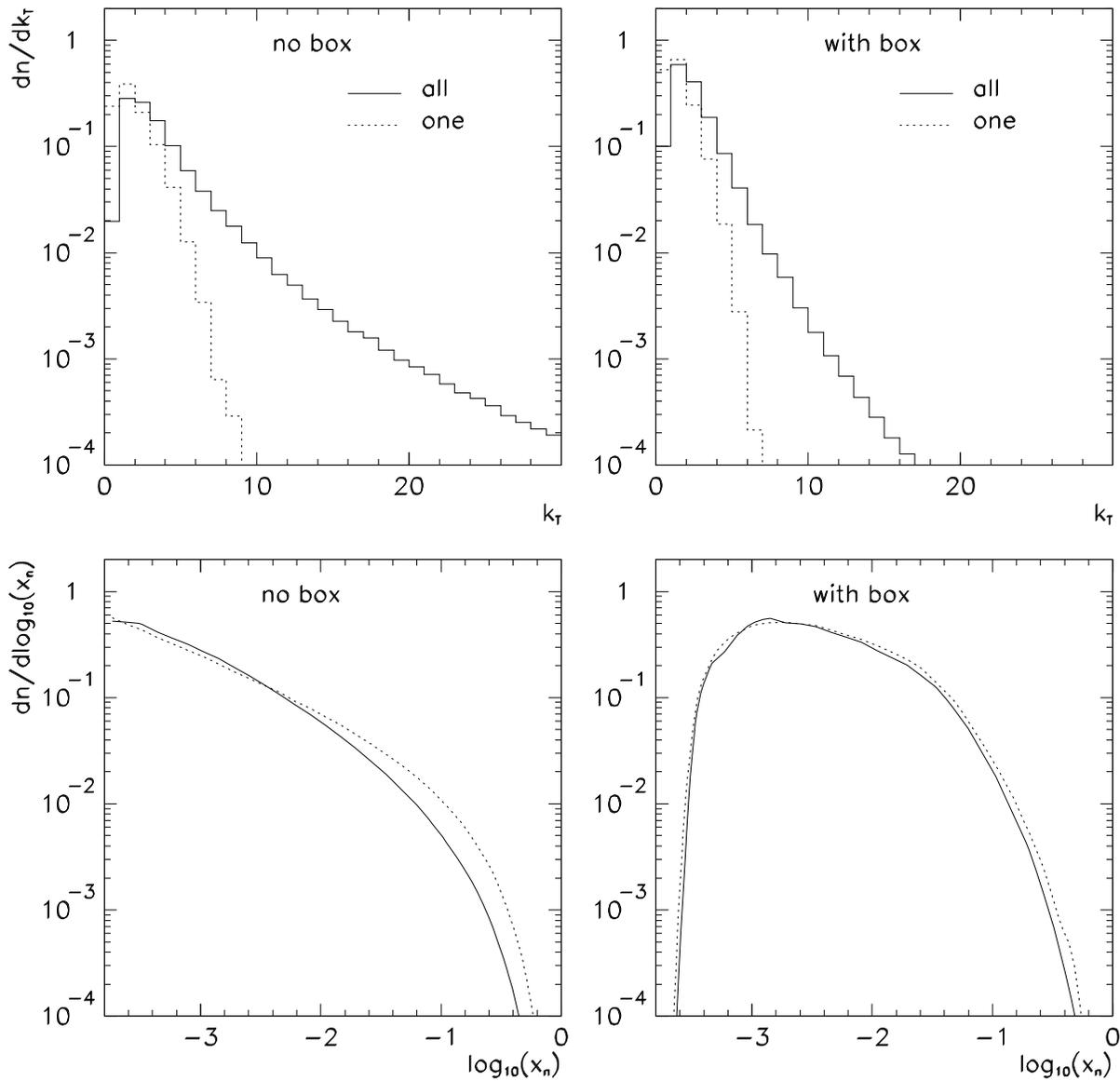,height=18cm,width=18cm}
               }
    \vspace*{-0.5cm}
     \caption{
     The transverse  $k_{\bot}$ 
     and longitudinal $x_n$ momentum distributions 
     of the last exchanged gluon for all gluonic events (``{\it no box}'')
     and the events with the quark pair produced (``{\it with box}'')
     in the all-loop (solid) and one-loop (dotted) approximations.
}\label{fig:5}
\end{figure}
\newpage

\begin{figure}[htb]
   \vspace*{-1cm}
    \centerline{
     \psfig{figure=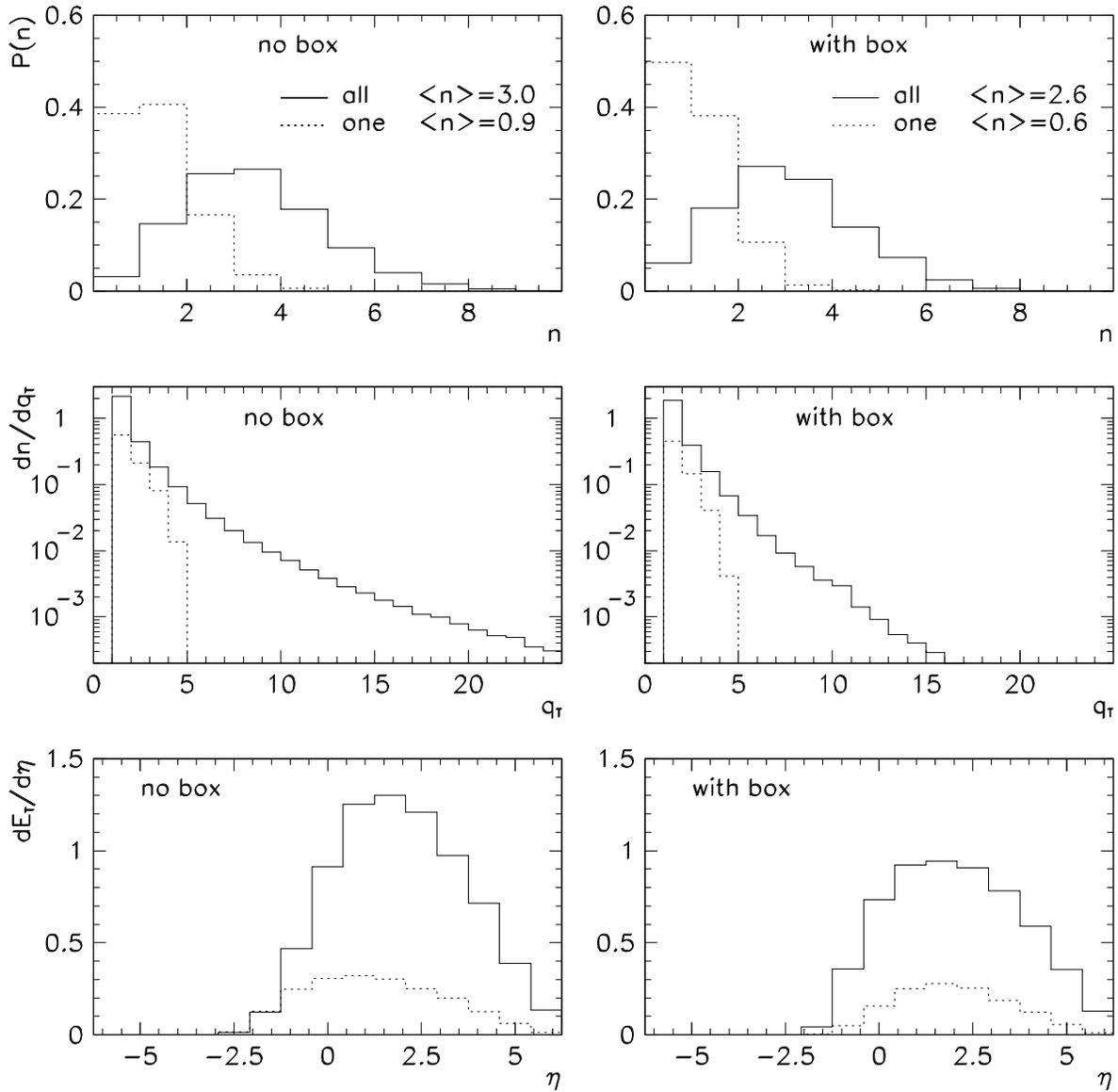,height=18cm,width=18cm}
               }
    \vspace*{-0.5cm}
     \caption{The final state gluon characteristics in the
     all-loop (solid) and one-loop (dotted) approximations.
     The influence of the quark pair production  is
     shown on the right (``{\it with box}'' plots).
}\label{fig:4}
\end{figure}
\newpage

\begin{figure}[htb]
   \vspace*{-1cm}
    \centerline{
     \psfig{figure=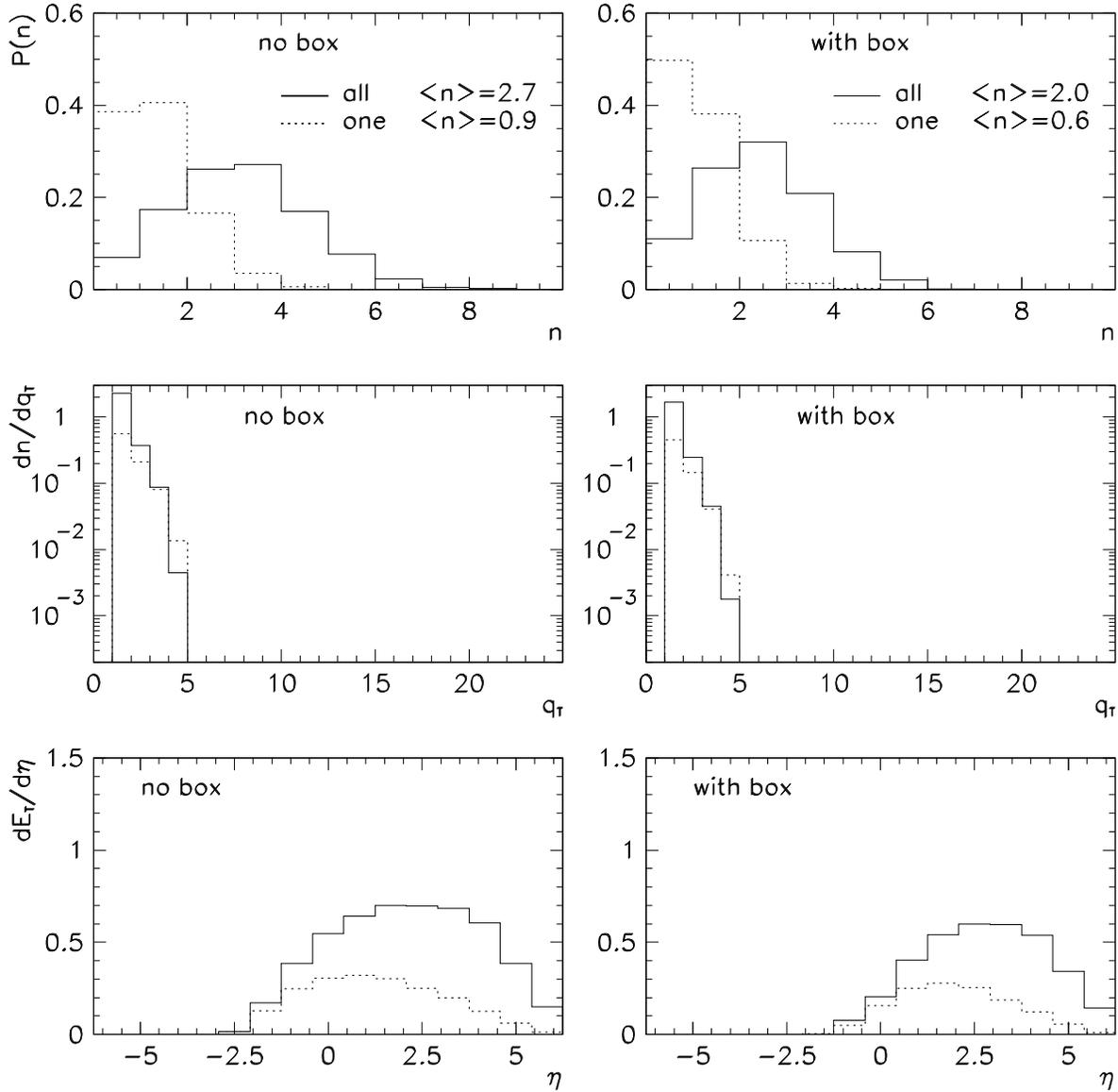,height=18cm,width=18cm}
               }
    \vspace*{-0.5cm}
     \caption{The same as Fig.~\ref{fig:4} but now the one-loop
     stopping condition is imposed in the all-loop generation.
}\label{fig:4stop}
\end{figure}
\newpage

\begin{figure}[htb]
   \vspace*{-1cm}
    \centerline{
     \psfig{figure=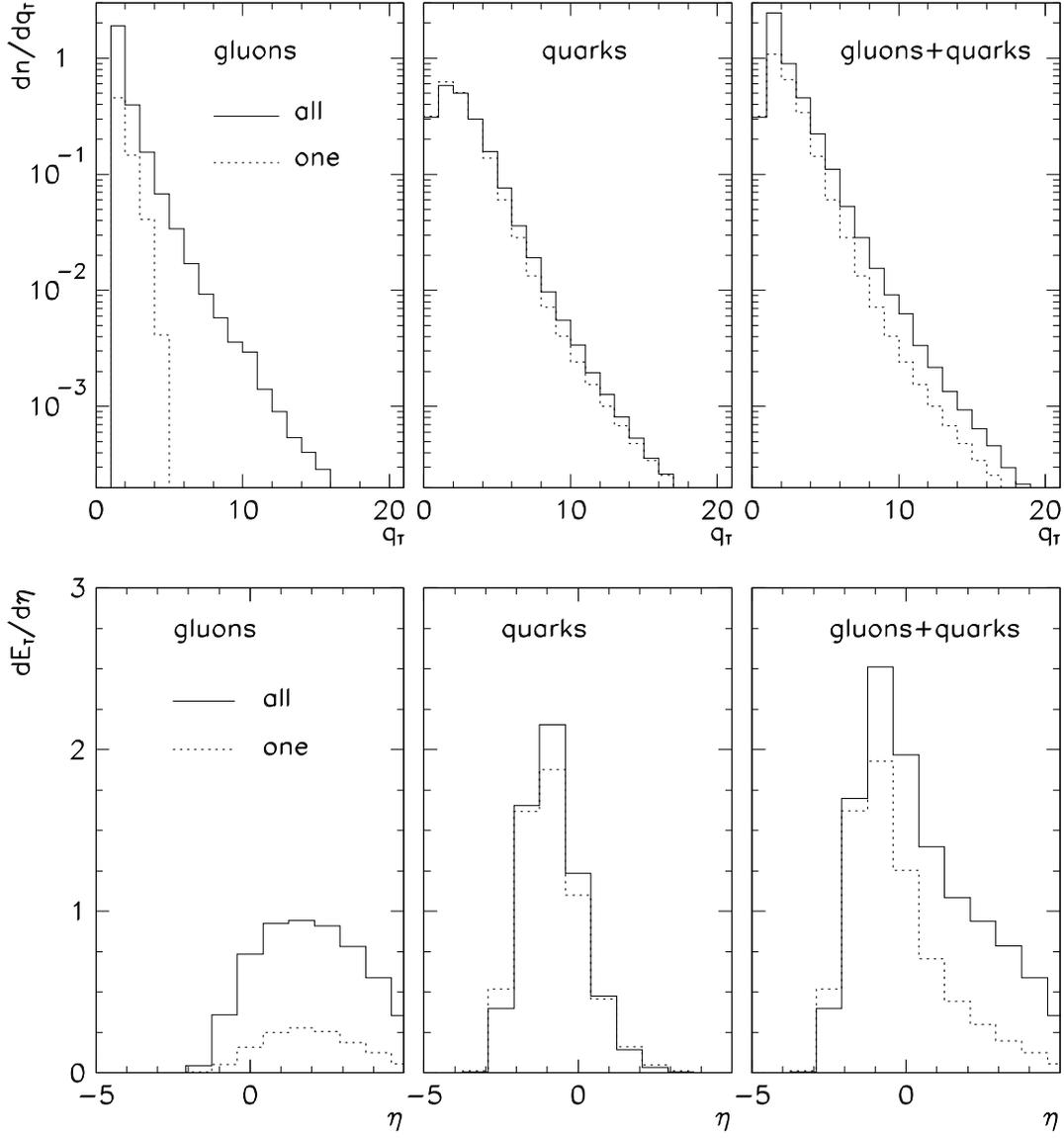,height=18cm,width=18cm}
               }
    \vspace*{-0.5cm}
     \caption{The $k_{\bot}$-spectrum and $E_{T}$ flow
     for final state partons.  Gluon and quark contributions
     are shown separately.
}\label{fig:6}
\end{figure}
\newpage

\begin{figure}[htb]
   \vspace*{-1cm}
    \centerline{
     \psfig{figure=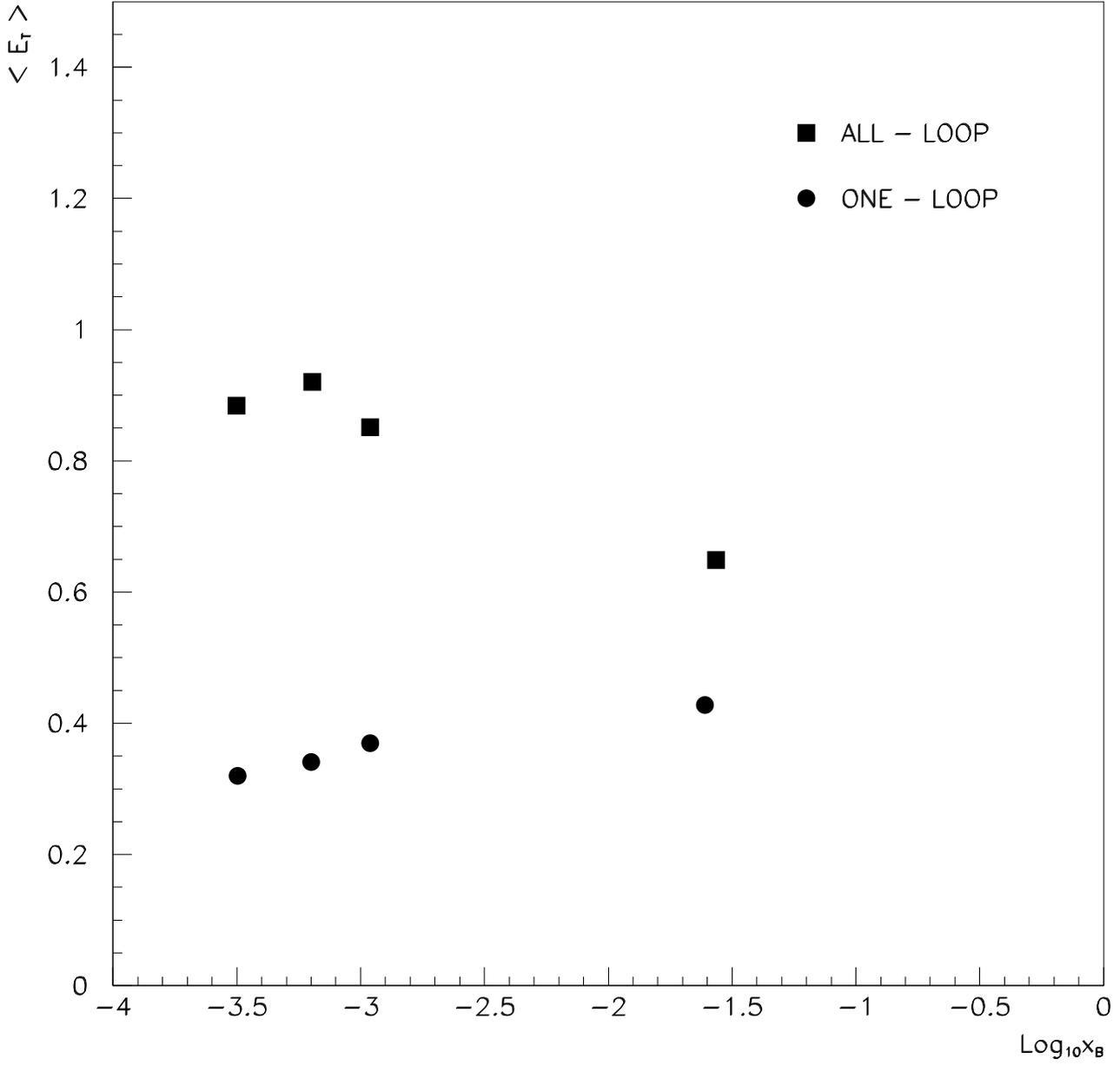,height=18cm,width=18cm}
               }
    \vspace*{-0.5cm}
     \caption{The average value of the transverse energy 
     in the central rapidity region $-0.5<\eta^{*}<0.5$
     in the $\gamma p$ CMS frame as a function
     of the Bjorken variable $x$.
}\label{fig:7}
\end{figure}
\newpage

\begin{figure}[htb]
   \vspace*{-1cm}
    \centerline{
     \psfig{figure=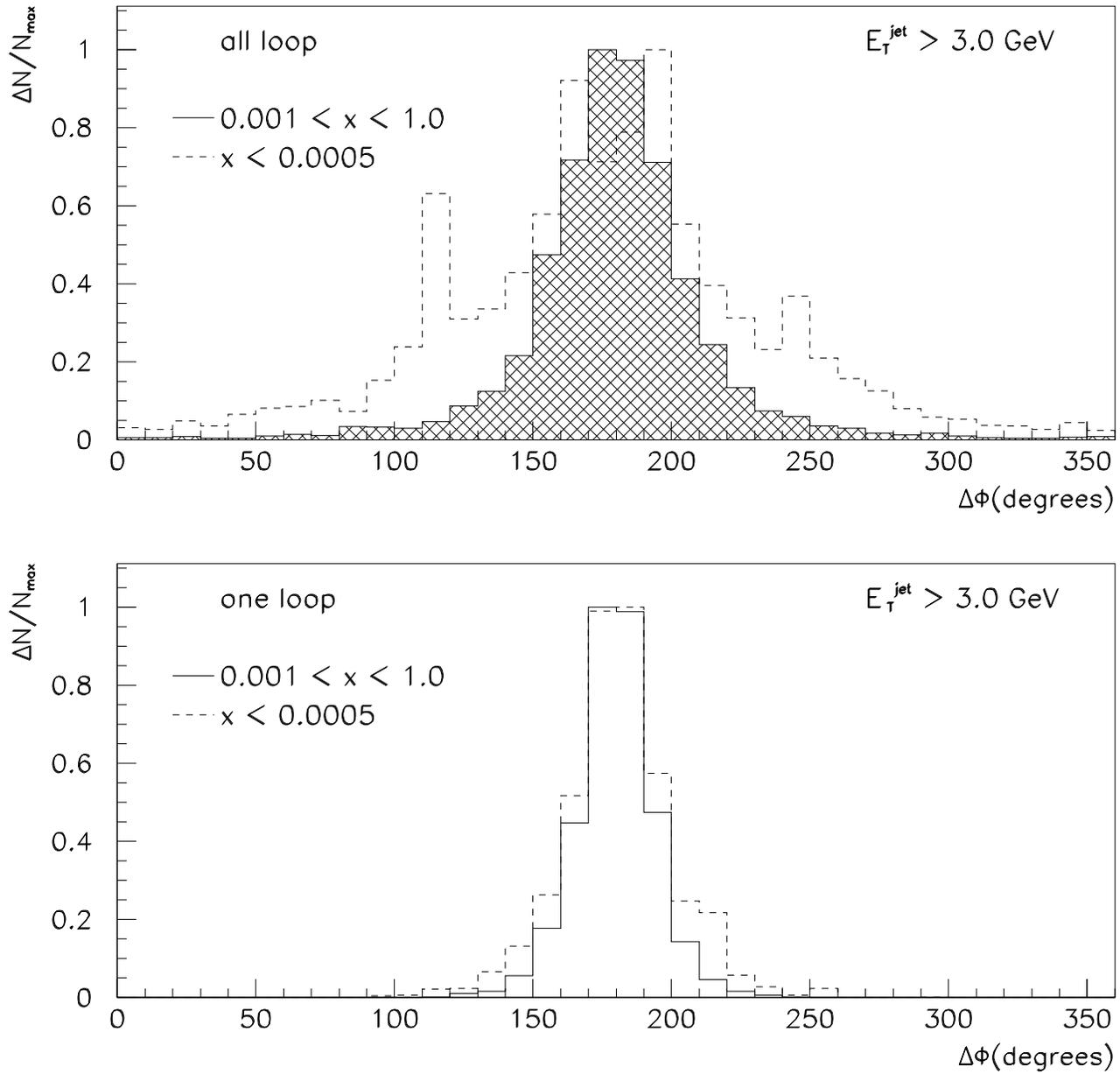,height=18cm,width=18cm}
               }
    \vspace*{-0.5cm}
     \caption{The difference between the azimuthal angles of  the quark
     and antiquark momenta 
     in the $\gamma p$ CMS frame for different values of the Bjorken
     variable $x$.
}\label{fig:8}
\end{figure}

\newpage

\begin{figure}[htb]
   \vspace*{-1cm}
    \centerline{
     \psfig{figure=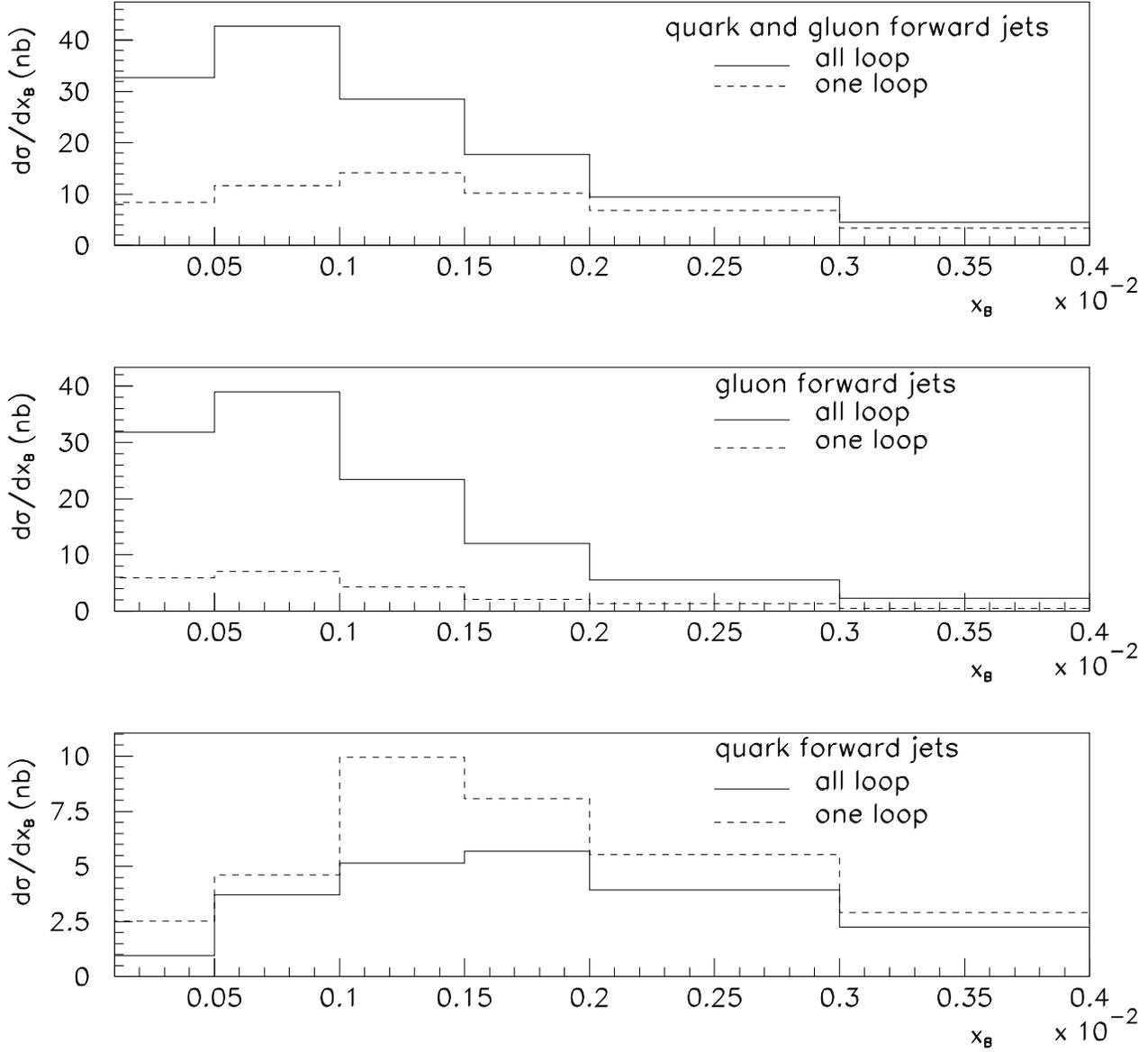,height=18cm,width=18cm}
               }
    \vspace*{-0.5cm}
     \caption{The forward jet cross section as a function of the
     Bjorken variable $x_B$, with the H1 analysis cuts. The gluon 
     and quark forward jet cross sections are shown on separate plots.    
}\label{fig:9}
\end{figure}

\end{document}